\documentclass[aps,prl,reprint,superscriptaddress,10pt]{revtex4-2}

\usepackage{graphicx,amsmath,amssymb,dcolumn,physics,color}

\graphicspath{{./figs/}{./}}
\newcommand{\sm}{SM}

\begin{document}

\title{Coarse-graining dynamics to maximize irreversibility}

\author{Qiwei Yu}
\affiliation{Lewis-Sigler Institute for Integrative Genomics, Princeton University, Princeton, NJ 08544}

\author{Matthew P.~Leighton}
\affiliation{Department of Physics and Quantitative Biology
Institute, Yale University, New Haven, CT 06511}

\author{Christopher W.~Lynn}
\email{Corresponding author: christopher.lynn@yale.edu}
\affiliation{Department of Physics and Quantitative Biology
Institute, Yale University, New Haven, CT 06511}
\affiliation{Wu Tsai Institute, Yale University, New Haven, CT 06510}

\begin{abstract}
    In many far-from-equilibrium biological systems, energy injected by irreversible processes at microscopic scales propagates to larger scales to fulfill important biological functions.
    But given dissipative dynamics at the microscale, how much irreversibility can persist at the macroscale?
    Here, we propose a model-free coarse-graining procedure that merges microscopic states to minimize the amount of lost irreversibility.
    Beginning with dynamical measurements, this procedure produces coarse-grained dynamics that retain as much information as possible about the underlying irreversibility.
    In synthetic and experimental data spanning molecular motors, biochemical oscillators, and recordings of neural activity, we derive simplified descriptions that capture the essential nonequilibrium processes.
    These results provide the tools to study the fundamental limits on the emergence of macroscopic irreversibility.
\end{abstract}

\maketitle

\emph{Introduction}---Biological systems are intrinsically nonequilibrium. The continuous dissipation of free energy is required to fulfill a wide range of biological functions~\cite{Gnesotto2018_brokenreview,Yang2021_physicalcells}, ranging from error correction~\cite{Hopfield1974_kineticspecificity,Ninio1975_kineticdiscrimination,Murugan2012_speedproofreading,Murugan2014_discriminatorysystems,Yu2022_thediscrimination} and environmental sensing~\cite{Ouldridge2017_thermodynamicssystems,Mehta2016_landauernetworks, TenWolde2016_fundamentalsensing, Govern2014_optimalsystems, Hathcock2023g_a,Hathcock2024_time-reversalsensing,pagare2023theoretical,Tjalma2023_trade-offsprediction} to communication \cite{bryant2023physical} and collective behaviors~\cite{Ferretti2022_signaturesflocking,Yu2022_energytransition,Ferretti2024_outsystems_a}.
Energy dissipation occurs at the microscopic scale through irreversible reaction cycles and then propagates across scales to achieve emergent biological functions. 
For example, in cytoskeletal networks, energy consumed by motor proteins at the molecular scale propagates to the mesoscale to power active flows of microtubule bundles~\cite{Foster2023_dissipationmaterial}.
Similarly, neurons and glia in the human brain account for up to 20\% of the body's metabolism to power neural activity that is responsible for higher-order cognition~\cite{harris2012synaptic, lynn2021broken, Lynn2022_decomposingsystems}.
Yet despite the clear generation of nonequilibrium dynamics at small scales, it remains unclear how much irreversibility can propagate to large scales.

A natural approach for connecting observables (such as irreversibility) across scales is the renormalization group (RG)~\cite{Kadanoff1966_scalingc,Wilson1975_theproblem}, which has provided key insights into active and living systems~\cite{Tu2023_thesystems, Meshulam2019_coarseneurons,Cavagna2019_dynamicalswarms,Cavagna2023_naturaldimensions,Yu2021_inversenetworks}.
To implement RG, one must choose an iterative coarse-graining procedure, which defines the macroscopic degrees of freedom.
Conventionally, coarse-graining is based on locality: Microscopic variables are combined when they are close in space~\cite{Kadanoff1966_scalingc, Wilson1975_theproblem}.
However, locality is less obvious in complex biological systems, where interactions can span long distances (as in networks of neurons) and states can be abstract without reference to physical proximity (as in chemical kinetics).
A key result in the study of nonequilibrium systems is that all coarse-grainings destroy information about the underlying dynamics, thus leading to (sometimes dramatic) reductions in irreversibility \cite{Esposito2012_stochasticgraining, Busiello2019_entropydynamics, Busiello2019_entropyloss, Bo2014_entropytime-scales, Cocconi2022_scalingmedia, Yu2021_inversenetworks,Yu2022_state-spacedimensions, Yu2024_dissipationscales}.
This means that in each step of RG, there is a unique coarse-graining with maximum irreversibility, which retains as much information as possible about the dissipative dynamics.

Here, we propose a coarse-graining procedure that minimizes the drop in irreversibility from microstates to macrostates.
The result is a model-free framework that identifies, at each level of coarse-graining, an effective description that captures as much of the underlying irreversibility as possible. 
Studying molecular motors, biochemical oscillators, and neural activity, we consistently find that a large amount of irreversibility can be preserved at high levels of description. Moreover, the coarse-grained dynamics capture key nonequilibrium features, such as limit cycles and directed flows, which highlight the large-scale functions in each system. 
These results pave the way for general applications to living systems far from equilibrium. To begin, we must define the maximum irreversibility coarse-graining.

\emph{Maximum irreversibility coarse-graining}---We consider a system with $N$ microscopic states and transition rates $k_{ij}$ from state $i$ to state $j$. The steady-state probabilities $P_i$ and fluxes $J_{ij}=k_{ij}P_i$ satisfy the flux balance condition $\sum_j J_{ij} = \sum_j J_{ji}$ for each state $i$. 
The total irreversibility of the system is
\begin{equation}
    \sigma = \frac{1}{2}\sum_{i,j} \hat{\sigma}(J_{ij}, J_{ji}),
\end{equation}
where the contribution from each pair of states is
\begin{equation}
    \hat{\sigma}(J_{ij}, J_{ji}) = (J_{ij}-J_{ji}) \ln \frac{J_{ij}}{J_{ji}},
\end{equation}
and the factor $\frac{1}{2}$ avoids double counting. Even if the dynamics are non-Markovian, as occurs generically after coarse-graining, we can still define $\sigma$; it measures the ``local" irreversibility, or the explicit violation of time-reversal symmetry in individual transitions~\cite{Lynn2022_emergencesystems, Lynn2022_decomposingsystems}.

Coarse-graining (CG) combines microstates ($i,j,\hdots$) into macrostates ($\alpha,\beta,\hdots$), leading to probabilities $P_\alpha = \sum_{i\in \alpha} P_i$ and fluxes $J_{\alpha\beta} = \sum_{i\in \alpha,j\in \beta} J_{ij}$.
Each choice of CG leads to a drop in irreversibility,
\begin{align}
    \Delta \sigma = \frac{1}{2}\qty[\sum_{i,j} \hat{\sigma}(J_{ij}, J_{ji})- \sum_{\alpha,\beta} \hat{\sigma}(J_{\alpha\beta}, J_{\beta\alpha})].
\end{align}
It is straightforward to show that $\Delta \sigma \ge 0$ [see Supplemental Material (\sm)], such that coarse-graining can only reduce the apparent irreversibility of the system  \cite{Esposito2012_stochasticgraining}.
This means that, for a given number of macrostates, there is a unique CG that minimizes $\Delta \sigma$, thus retaining as much information as possible about the dissipative dynamics.
However, because the number of CGs explodes combinatorially with $N$, searching for the optimal CG with maximum irreversibility is infeasible.

Instead, following RG, we can construct macrostates by iteratively combining pairs of microstates. At each step, we search over each pair of states $\{\alpha, \beta\}$ and compute the drop in irreversibility due to their merger,
\begin{align}
\label{eq_pair}
    \Delta\sigma_{\alpha\beta} &= \hat{\sigma}(J_{\alpha\beta}, J_{\beta\alpha}) +\sum_{\gamma\neq\alpha,\beta} \Big[\hat{\sigma}(J_{\alpha\gamma}, J_{\gamma\alpha}) + \hat{\sigma}(J_{\beta\gamma}, J_{\gamma\beta}) \nonumber \\
    &\quad\quad\quad - \hat{\sigma}(J_{\alpha\gamma}+J_{\beta\gamma}, J_{\gamma\alpha}+J_{\gamma\beta}) \Big].
\end{align}
Combining the pair of states that minimizes $\Delta\sigma_{\alpha\beta}$ and repeating this process until the desired number of macrostates is reached, 
we arrive at an efficient CG that locally maximizes the irreversibility.

The form of Eq.~(\ref{eq_pair}) immediately hints at the structure of optimal CGs.
To minimize $\Delta\sigma_{\alpha\beta}$, one seeks to combine states $\{\alpha,\beta\}$ with not only low internal irreversibility $\hat{\sigma}(J_{\alpha\beta}, J_{\beta\alpha})$, but also low irreversibility due to fluxes with neighbors $\gamma$ in the system $\hat{\sigma}(J_{\alpha\gamma}, J_{\gamma\alpha}) + \hat{\sigma}(J_{\beta\gamma}, J_{\gamma\beta})$. Additionally, once combined, the optimal states should have fluxes that align to produce high irreversibility with the rest of the system $\hat{\sigma}(J_{\alpha\gamma}+J_{\beta\gamma}, J_{\gamma\alpha}+J_{\gamma\beta})$. We note, however, that even with the tools to construct the optimal CG, there is no guarantee that a small number of states can capture a large amount of irreversibility. Whether a good CG exists will depend critically on the system itself.

\begin{figure}[t]
    \centering
    \includegraphics[width=\linewidth]{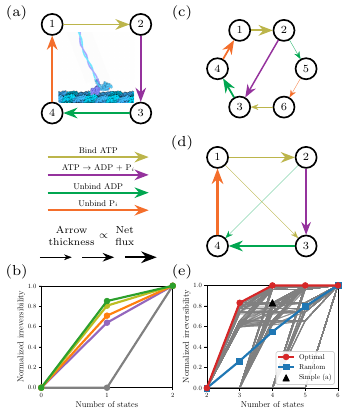}
    \caption{Coarse-graining the kinesin chemical reaction cycle.
    (a)~Minimal four-state model of the reaction cycle as the motor steps along the microtubule~\cite{Fisher2001_simple}. Colors and thicknesses of arrows represent the types and strengths of net fluxes [same for (c--d)]. Inset image adapted from PDB~\cite{berman2000protein,goodsell2020insights}.
    (b)~Fraction of irreversibility preserved under coarse-graining for the four-state model. Colors correspond to coarse-graining the different steps in (a); grey curves correspond to diagonal mergers (combining states 1 and 3 or 2 and 4), which destroy the irreversibility.
    (c)~Six-state model with separate cycles for forward and backward steps~\cite{liepelt2007kinesin}. 
    (d)~Optimal four-state CG of the six-state model.
    (e)~Fraction of irreversibility preserved under coarse-graining for the six-state model. Grey curves show all possible CGs, red is optimal, blue is random (averaging over all CGs), and black is the four-state model in (a). See \sm\ for details and parameters. 
    }
    \label{fig:kinesin}
\end{figure}

\begin{figure*}[t]
    \centering
    \includegraphics[width=\linewidth]{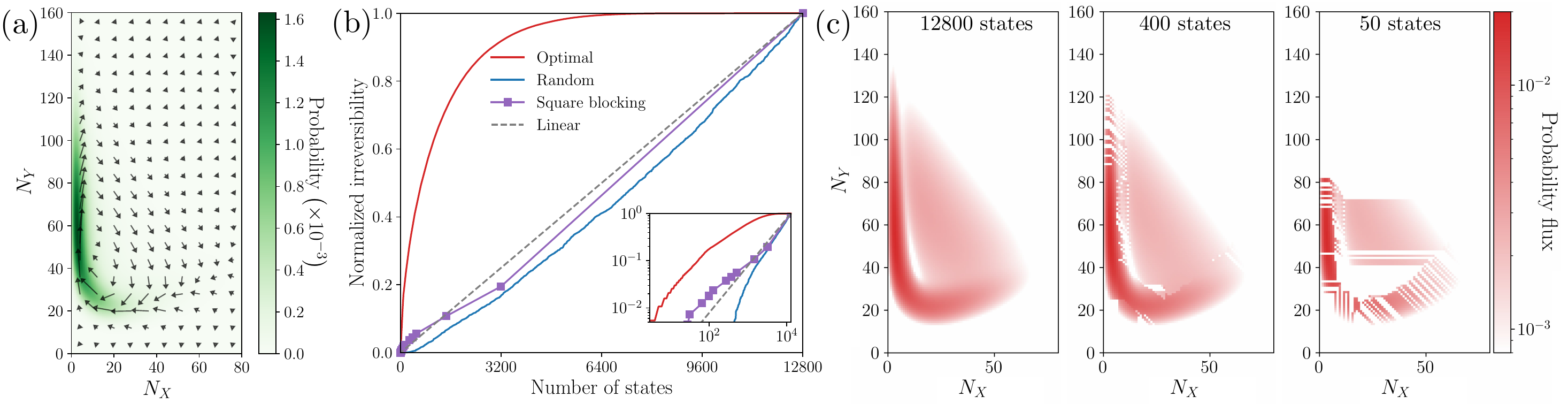}
    \caption{
        Coarse-graining biochemical oscillations.
        (a) Steady-state probability density (colors) and net fluxes (arrows) in the Brusselator model of biochemical oscillations~\cite{Nicolis1977, Fritz2020_stochasticbrusselator}. 
        Fluxes are 
        averaged over local neighborhoods.
        (b) Fraction of irreversibility preserved after coarse-graining for three different procedures: optimal (red), random (blue), and square blocking (purple). Inset displays the same data on log-log axes.
        (c) Net fluxes between states in the full system with $N = 12800$ states (left) and under optimal CGs with $400$ (center) and $50$ (right) macrostates. Within each macrostate, fluxes are lost due to coarse-graining. 
        See \sm\ for details and parameters. 
    }
    \label{fig:brusselator}
\end{figure*}

\emph{Molecular motor}---To illustrate this procedure, we consider the chemical kinetics of the molecular motor kinesin-1, one of a family of motor proteins that transport cargo along microtubules within cells~\cite{Hirokawa2009_kinesintransport}. 
The stepping of the motor can be modeled as a four-state reaction cycle [Fig.~\ref{fig:kinesin}(a)] with net fluxes representing the binding, unbinding, and hydrolysis of ATP (and its reaction products)~\cite{Fisher2001_simple}. Here, one full forward (reverse) chemical cycle corresponds to one forward (reverse) step of the motor along the microtubule.
In this case, the full irreversibility is equivalent to the microscopic entropy production of the reaction dynamics.
There are six different ways to coarse-grain the model, one for each pair of states.
Two CGs combine states diagonally (1 and 3 or 2 and 4), destroying the reaction cycle and reducing the irreversibility to zero [Fig.~\ref{fig:kinesin}(b), grey lines].
By contrast, the four remaining CGs combine adjacent states, removing one step from the reaction cycle but preserving the overall flux.
Even with the same net flux, each of these CGs yields a distinct drop in irreversibility [Fig 1(b), colored lines].
Among these options, we find that coarse-graining the ADP unbinding step minimizes the loss of irreversibility.

A more detailed model of kinesin incorporates separate cycles for the forward and reverse steps of the motor [Fig.~\ref{fig:kinesin}(c)]~\cite{liepelt2007kinesin}.
Just by expanding to six states, there are now 65 different CGs with four states and 90 with three states. Yet across each level of description, our iterative algorithm identifies the globally optimal CG with maximum irreversibility [Fig.~\ref{fig:kinesin}(e)], which can be computed exactly for this small network by enumerating all possible CGs.
These optimal CGs retain much more of the underlying nonequilibrium dynamics than random combinations of microstates.
In fact, with only four states, the optimal CG retains all of the microscopic irreversibility [Fig.~\ref{fig:kinesin}(d)]. This reduced description contains the same dominant reaction cycle as the original four-state model in Fig.~\ref{fig:kinesin}(a), but also preserves additional irreversibility by incorporating secondary cycles from the six-state model [Fig.~\ref{fig:kinesin}(c)]. 
Furthermore, the optimal three-state CG preserves 83\% of the total irreversibility, approximately the same amount as the original four-state model [Fig.~\ref{fig:kinesin}(e), black triangle].
These results demonstrate how, even in a simple system, one can achieve reduced descriptions with coarse-grained states and fluxes that still maintain the key nonequilibrium dynamics.

\emph{Biochemical oscillator}---The efficiency of our CG procedure opens the door for applications to systems with many more microstates.
Consider the Brusselator model for biochemical oscillators~\cite{Nicolis1977,Fritz2020_stochasticbrusselator}, which contains $N\sim 10^4$ states spanned by the numbers $N_X$ and $N_Y$ of two molecules $X$ and $Y$ (see \sm\ for details).
Chemical reactions involving the two molecules induce net fluxes between neighboring states [Fig.~\ref{fig:brusselator}(a)]. As in many biochemical oscillations, these fluxes combine to form a macroscopic loop; this suggests that, by focusing on the central loop of flux, a coarse-grained description may be able to capture much of the nonequilibrium dynamics. 

To preserve locality, we only consider CGs that combine neighboring states.
If states are combined at random, then the nonequilibrium dynamics are quickly lost to coarse-graining, and we observe a super-linear drop in irreversibility [Fig.~\ref{fig:brusselator}(b), blue]. For comparison, following RG, we can recursively combine neighboring states into square blocks~\cite{Kadanoff1966_scalingc,Yu2021_inversenetworks}. This approach only retains a slightly larger fraction of the irreversibility than random [Fig.~\ref{fig:brusselator}(b), purple]. 
Strikingly, the optimal CG, computed using our iterative procedure, preserves orders of magnitude more irreversibility than these na\"{i}ve methods [Fig.~\ref{fig:brusselator}(b), red]. 

To understand how the optimal CG preserves irreversibility, we can investigate the structure of the inferred macrostates. In Fig.~\ref{fig:brusselator}(c), we illustrate the strengths of the net fluxes between states at different levels of coarse-graining; note that these fluxes also indicate the boundaries between states since the fluxes within macrostates are destroyed. 
We see that the optimal procedure partitions the state space along the limit cycle, combining states perpendicular to the cycle in order to maintain as much flux as possible [Fig.~\ref{fig:brusselator}(c), center].
As the number of states decreases, the partition becomes coarser, but the limit cycle remains the dominant feature [Fig.~\ref{fig:brusselator}(c), right].
For comparison, if we combine random states or blocks of adjacent states, then the fluxes are almost entirely lost (see \sm).
We therefore find that a large amount of irreversibility in biochemical reactions can be compressed into a small number of states. Moreover, simply by preserving the dissipative dynamics (without any knowledge of the underlying physics), the optimal CG uncovers the central dissipative structure in the system.

\begin{figure*}[t]
    \centering
    \includegraphics[width=0.75\linewidth]{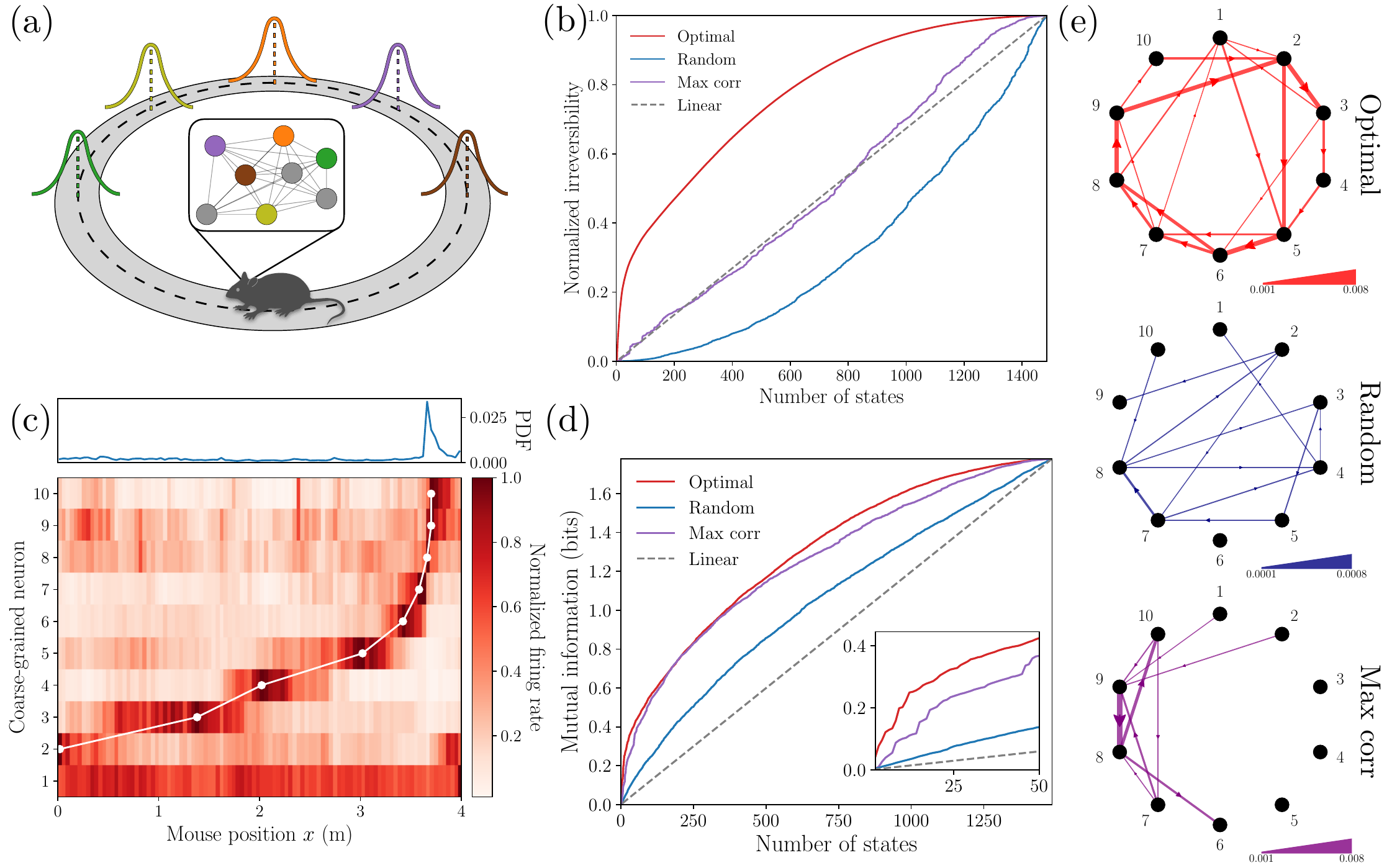}
    \caption{
        Coarse-graining neural activity in the hippocampus.
        (a) A mouse runs along a one-dimensional virtual track as the activities of $N = 1485$ neurons are recorded in the hippocampus~\cite{Gauthier2018_ahippocampus}. Approximately $30\%$ of the neurons are place cells (indicated by colors in the neural network), which fire preferentially at specific locations along the track (indicated by location-dependent firing rates known as place fields). Mouse graphic adapted from Ref.~\cite{long_li_2021_5496316} (CC BY 4.0).
        (b) Fraction of irreversibility preserved after coarse-graining for three different procedures: optimal (red), random (blue), and combining maximally correlated neurons (purple). 
        (c) Place fields of 10 macrostates computed using the optimal CG with white points indicating the peaks. 
        Top inset represents the distribution of mouse positions along the track.
        (d) Mutual information between coarse-grained neural activity and mouse position for different CG procedures. Inset is a zoomed-in view for the coarsest scales. 
        (e) Probability flux between 10 macrostates. Note that the scale for random is 10 times smaller than optimal and max-corr. 
    }
    \label{fig:hippocampus}
\end{figure*}

\emph{Neural activity}---Thus far, we have focused on model systems in which the microscopic details are fully known and the macroscopic irreversibility is self-evident.
But what can coarse-graining reveal about real biological measurements, where the nonequilibrium dynamics are unknown?
To answer this question, we study recordings of neural activity in the hippocampus of a mouse~\cite{Gauthier2018_ahippocampus}.
Specifically, we consider the activity of $N = 1485$ neurons, recorded using two-photon calcium imaging as the mouse runs along a one-dimensional virtual track [Fig.~\ref{fig:hippocampus}(a); see \sm\ for details].
Approximately 30\% of these neurons are place cells, which preferentially fire when the mouse is in a specific location~\cite{OKeefe1976_placerat, OKeefe1979_thehippocampus, meshulam2017collective}. 
In this way, the hippocampus is thought to play a key role in encoding the animal's position and mapping features in its environment~\cite{blum1996model,Hazon2022_noiserepresentations,Nardin2023_theexperience}.

As the mouse moves forward along the track, it generates an irreversible cycle of flux in physical space.
Yet it remains unclear whether this physical flux induces an irreversible flux in the abstract space of neural activity.
Here, we define the state of the population based on the most recent neuron to fire, and we count a transition $i\to j$ if neuron $i$ firing leads to neuron $j$ firing after a time delay $\Delta t$ (see \sm\ for details). We use a time delay of $\Delta t = 3$s, which we find produces neural dynamics with the largest irreversibility; we confirm that other choices for $\Delta t$ yield the same results, qualitatively (see \sm).
Thus, we arrive at a system with $N$ states, one for each neuron, where the fluxes $J_{ij}$ represent the flow of neural activity from neuron $i$ to neuron $j$.

For each coarse-grained description, merging states is equivalent to combining neurons, such that macrostates directly reflect the collective activity of subsets of the population. Without any knowledge of the system, one can combine neurons at random; in this case, fluxes in activity are quickly destroyed, leading to a rapid drop in irreversibility [Fig.~\ref{fig:hippocampus}(b), blue]. 
With knowledge of only the correlations between neurons, the most natural CG iteratively combines pairs of cells whose activities are most correlated (see \sm);
this method still only preserves an amount of irreversibility that is linear in the number of states [Fig.~\ref{fig:hippocampus}(b), purple]. Finally, with knowledge of the neural dynamics, our optimal CG uncovers groups of neurons that preserve macroscopic fluxes in activity, thus capturing much more of the underlying irreversibility [Fig.~\ref{fig:hippocampus}(b), red].

Visualizing the optimal CG, we find that it groups neurons into macroscopic ``place cells", which activate when the mouse is in a specific location [Fig.~\ref{fig:hippocampus}(c)]. Each macro-cell is characterized by a macroscopic ``place field", which defines its activity as a function of position. These place fields are not distributed uniformly along the track; instead, the spatial resolution increases in locations where the mouse spends more time [Fig.~\ref{fig:hippocampus}(c), top]. The lone exception is the first macro-cell, which contains the majority of the non-place cells in the population and therefore exhibits uniform activity along the track (see \sm). Thus, without any access to positional information, the optimal CG appears to learn an efficient representation of the mouse location. To confirm this intuition, we compute the mutual information between the coarse-grained activity and mouse position [Fig.~\ref{fig:hippocampus}(d)]; across all levels of description, the optimal CG achieves high spatial information.

Given the spatial structure of the optimal CG, as the mouse runs along the track, we should observe a directed cycle of transitions from one macrostate to the next. Indeed, for the ten-state CG in Fig.~\ref{fig:hippocampus}(c), the neural dynamics display a clear loop of flux that reflects the physical trajectory of the mouse [Fig.~\ref{fig:hippocampus}(e), top]. By contrast, while random CGs exhibit fluxes between macrostates, these fluxes are much smaller in magnitude and do not combine into a coherent cycle [Fig.~\ref{fig:hippocampus}(e), center]. Meanwhile, combining the most correlated neurons produces macrostates with high spatial information [Fig.~\ref{fig:hippocampus}(d)], but even this description fails to produce a large-scale cycle [Fig.~\ref{fig:hippocampus}(e), bottom]. Thus, by maximizing the irreversibility of neural dynamics (without any knowledge of the animal's behavior), we arrive at a reduced description that uncovers the main driver of large-scale fluxes in the hippocampus: spatial navigation.

\emph{Discussion}---In living systems, irreversible dynamics are driven at microscopic scales by dissipative processes that consume energy. These nonequilibrium dynamics then propagate across spatiotemporal scales to power biological functions that are critical for life. Bridging microscopic irreversibility with macroscopic biological function has thus remained a long-standing goal in the physics of living systems~\cite{schrodinger1948life}.

Here, we develop a model-free procedure for constructing optimal coarse-grained descriptions of irreversible systems. Our framework hinges on one key insight: Local irreversibility of a system can only decrease with coarse-graining~\cite{Esposito2012_stochasticgraining}. 
Thus, for a desired number of macrostates, there is generically a unique CG that retains as much information as possible about the underlying nonequilibrium dynamics. Taking inspiration from RG, we propose an iterative procedure that combines pairs of states to locally maximize the irreversibility of the coarse-grained description. 
Thus, in contrast to inferring microscopic irreversibility from coarse-grained measurements~\cite{li2019quantifying,otsubo2020estimating,horowitz2020thermodynamic,Cisneros2023_dissipativeirreversibility,ghosal2022inferring,Skinner2021_improvedsystems,Skinner2021_estimatingdistributions,vanDerMeer2022_thermodynamicdistributions,Harunari2022_what,leighton2024jensen,Li2024_measuringpatterns}, we have taken a complementary approach by investigating how irreversibility can be used to infer simplified representations of nonequilibrium systems. 
Applying our framework to molecular motors, biochemical oscillators, and recordings of neural activity, we consistently find that a large amount of irreversibility can be compressed into a small number of macrostates. Moreover, in each instance, our procedure uncovers coarse-grained dynamics that capture the essential nonequilibrium processes.

The proposed framework is general, applying (without model assumptions) to any dynamical time-series data. It can therefore immediately be used to investigate the nature of nonequilibrium dynamics spanning a range of complex living systems, including cytoskeletal networks~\cite{Foster2023_dissipationmaterial}, cellular signaling~\cite{bryant2023physical}, whole-brain neural dynamics~\cite{lynn2021broken}, and animal behavior~\cite{Berman2014_mappingflies, Costa2024_ascales, castellano2000nonequilibrium}. Across these distinct contexts, do there exist reduced descriptions that capture large amounts of irreversibility? And if so, are there consistent dissipative structures at small scales that support irreversible dynamics at large scales? The framework presented here provides the tools to begin answering these questions.

\emph{Acknowledgements}---This work was supported in part by a Harold W. Dodds Fellowship (Q.Y.), a Mossman Postdoctoral Fellowship and a Seessel Postdoctoral Fellowship (M.P.L.), and by support from the Department of Physics and the Quantitative Biology Institute at Yale University (C.W.L).

\bibliography{cg}

\end{document}